\input epsf
\documentstyle[12pt]{article}
\newlength{\dinwidth}
\newlength{\dinmargin}
\oddsidemargin  - 1.0in
\newcommand{\Brbsg}{${\rm BR}(B\to X_s\gamma)$ }

\newcommand{\bdsee}{$B\to X_{d(s)}e^+e^ - $}

\newcommand{\be}{\begin{equation}}
\newcommand{\ee}{\end{equation}}
\newcommand{\bea}{\begin{eqnarray}}
\newcommand{\eea}{\end{eqnarray}}
\renewcommand{\thefootnote}{\fnsymbol{footnote}}

\begin{document}

\begin{titlepage}
\setcounter{page}{0}
\rightline{YerPhI - 1554(3) - 2000}
\rightline{NRCPS-HE-2000-21}
\vspace{2cm}
\begin{center}
{\Large{\bf Direct CP-asymmetry in Inclusive Rare B-decays in 2HDM.}}
\end{center}
\vspace{1cm}
\begin{center}
{\large{\it H. H. Asatryan, 
H. M. Asatrian\footnote{{\it hrachia@moon.yerphi.am}}, G. K. Yeghiyan}}  \\
\vspace{0.5cm}
{\it Yerevan Physics Institute, \\
2 Alichanian Br.St. Yerevan, Armenia}\\
\vspace{0.5cm}
{\large{\it G.K.Savvidy}}  \\
\vspace{0.5cm}
{\it National Research Center Demokritos, \\
Agia Paraskevi, GR-15310  Athens, Greece}
\end{center}
\vspace{0.5cm}

\vspace{2cm}
\centerline{{\bf Abstract}}
The direct CP-asymmetry in the inclusive $B \to X_d \gamma$ and
$B \to X_d e^+ e^ - $  decays
is investigated in the two-Higgs doublet extension of
the Standard Model (2HDM). The investigation is performed in the lowest
non-vanishing order of the perturbation theory using the existing 
restrictions on the 2HDM parameters space. It is shown that the direct 
CP-asymmetry in the
$B \to X_d \gamma$ decay can deviate significantly from the Standard
Model predictions. In the presence of only one source of CP-violation
(the CKM matrix weak phase) $a_{CP}(B \to X_d \gamma)$ can have the sign
opposite to that in the SM. The new source of CP-violation can make
$|a_{CP}(B \to X_d \gamma)|$ arbitrary small (unlike the SM case)
and hence unmeasurable. Quantitatively, the obtained results suffer from
the uncertainty of the choice of renormalization scale. As for the
$B \to X_d e^+ e^ - $ rate asymmetry, its renormalization scale dependence
in the lowest non-vanishing order does not allow to conclude if this
quantity is efficient for testing New Physics beyond the Standard Model.

\end{titlepage}

\section{Introduction}
\setcounter{equation}{0}
\renewcommand{\theequation}{1.\arabic{equation}}
\renewcommand{\thefootnote}{\arabic{footnote}}

The experimental evidence for the Standard Model Higgs boson \cite{higgs}
increases the interest to the Higgs sector of the 
electroweak theory and in particular to the Two-Higgs doublet 
extension of the  Standard Model. 
The study of the rare B-decays is known to be an important source
of information for such an extension of the Standard Model and 
on the physics at the scales up to several TeV.
In particular these decays combined with the studies of  other 
CP-violating effects in K-  and B-physics can be useful for understanding
the CP-breaking phenomenon.

The first experimental evidence for the exclusive
rare decay channel $B\to K^*\gamma$ has been reported by the CLEO
collaboration \cite{28}. It was followed by the observation of the
inclusive decay $B\to X_s\gamma$ \cite{29}. At the present the experimental 
prediction for the $B\to X_s\gamma$ branching ratio from CLEO is  \cite{30}
\be
\label{exp}
<{\rm BR}(B \to X_s \gamma)> = \frac{1}{2}\left({\rm BR}(B \to X_s \gamma) +
{\rm BR}(\bar{B} \to \bar{X}_s \gamma)\right) = (3.15 \pm 0.54) \times 10^{ - 4}
\ee
For the other rare B-decays, such as $B\to X_d\gamma$, 
$B\to X_{s(d)}\ell^+\ell^ - $ etc., only 
upper experimental bounds exist. Hopefully these decays will be
observed at the B-factories which have already started operating.

Huge theoretical efforts have been devoted to the investigation
of the rare inclusive decays in the Standard Model (SM) and its extensions.
The most thorough study has been carried out for the $B\to X_s\gamma$
decay \cite{31}-\cite{13}. In particular, in \cite{18}-\cite{14} the NLO
corrections have been calculated in the SM reducing the 
uncertainty in theoretical prediction for
\Brbsg to the level of a few percent.
In the extensions of the SM the NLO corrections have been
considered in \cite{12}-\cite{51}. In these models the
theoretical uncertainty due to renormalization scale dependence
is still significant. The theoretical prediction for \Brbsg in the SM is in 
reasonable agreement with (\ref{exp}). Consequently
strong constraints on the parameters of the SM
extensions can be derived. In particular such constraints have been
obtained  for the Two-Higgs Doublet Model (2HDM) \cite{12,3}
and the Minimal Supersymmetric Standard Model (MSSM) \cite{47}-\cite{51}.

Another popular rare B-decay mode, $B\to X_s \ell^+\ell^ - $, has
been considered in the context of the SM and its extensions in \cite{15,34}, 
\cite{52}-\cite{48}. It is expected that this decay mode will be also
useful for testing the New Physics.

The direct CP-asymmetry in rare B-decays is another important source
of information on the physics beyond the SM. 
In the SM the only source of CP-violation is the single phase in the 
CKM-matrix. 
Due to known hierarchy of the CKM-elements, $b\to s$ transitions
have larger branching fractions while in $b\to d$ transitions
large CP-violating effects are expected.
Numerically (in the SM) the branching ratio for the 
$B\to X_d\gamma$ is $10\div20$
times smaller than for $B\to X_s \gamma$ but the CP-asymmetry of
the latter can reach 27\% 
to be compared with the prediction $ -(0.4\div1)\%$ \cite{4}
for the former. As it was pointed out in \cite{57},
the $B\to X_d\gamma $ rate asymmetry
is statistically more accessible than that for $B\to X_s\gamma$.
A similar hierarchy  between  the branching ratios and 
asymmetries holds (in the SM) also for the decays $B\to X_s e^+e^ - $ and
$B\to X_d e^+e^ - $ \cite{48}.

The aim of this paper is to perform a detailed study of the direct
CP-asymmetry in the inclusive $B\to X_d\gamma$ and 
$B\to X_d e^+e^ - $ decays in the lowest non-vanishing order
of the perturbation theory within the general 2HDM  -  the
simplest extension of the SM where another source of CP-violation
is present. 
The $b \to d \gamma$ transition asymmetry in the 
2HDM has been examined in \cite{10}.
In the present paper we improve the results of ref. \cite{10},
taking into account new theoretical results and experimental data. In 
particular the scale parameter dependence is investigated in details.
The included effects lead to a significant changes in the predictions
for the $B\to X_d\gamma$ decay rate asymmetry: 
it reaches -25\% when being negative and 37\% when being positive. 
The CP-asymmetry in $B \to X_d e^+ e^ - $ decay in the
SM and 2HDM has been studied in \cite{59,48} and \cite{49} respectively.
In ref. \cite{49} it was obtained that for a specific choice of the
parameters of the theory, $B \to X_d e^+ e^ - $ decay asymmetry
is more efficient for testing the physics beyond the SM
than the branching ratio. In present paper we show that when 
taking into account the renormalization scale
dependence in the lowest non-vanishing order
the  allowed interval of the 
CP-asymmetry for $B \to X_d e^+ e^-$ decay in the SM and 2HDM  does 
not differ
significantly. Therefore to clarify whether the  CP-asymmetry
can  be useful for testing the validity of the Standard Model
one should include  higher-order corrections to
$a_{CP}(B \to X_d e^+ e^ - )$ which have not been calculated yet.

The paper is organized as follows. 
In Section 2 we briefly describe the 2HDM, give the  formulae which
are needed for the
calculation of $B \to X_d \gamma$ and $B \to X_d e^+ e^ - $
rate asymmetries and discuss  the existing 
restrictions on the parameters of the model.
In Sections~3~and~4 we present the numerical results
for $a_{CP}(B \to X_d \gamma)$ and
$a_{CP}(B \to X_d e^+ e^ - )$ respectively. Finally Section~5 is
devoted to the summary of the obtained results and conclusions.

\section{Formulae for $B \to X_{d} \gamma$, $B \to X_{d} e^+ e^ - $
Rate Asymmetries and Constraints on Parameters of the Model.}
\subsection{The Two-Higgs Doublet Model.}
\setcounter{equation}{0}
\renewcommand{\theequation}{2.1.\arabic{equation}}
\renewcommand{\thetable}{2.1.\arabic{table}}
It is known that in the SM the rare decays
$B\to X_{d(s)}\gamma$ and \bdsee are loop-induced:
in the lowest order they proceed via exchange of up-type quarks
and $W^{\pm}$ bosons in the loops. In the 2HDM there are 
additional diagrams with W-bosons replaced by the charged
Higgs bosons ($H^{\pm}$). The Yukawa interaction of  the quarks
with the Higgs doublets $\Phi_1$ and $\Phi_2$ in general is \cite{11}
\be
{\cal L} =\bar{Q}_L(\Gamma_1^d\Phi_1+\Gamma_2^d\Phi_2)d_R+
\bar{Q}_L(\Gamma_1^u\widetilde{\Phi}_1+\Gamma_2^u\widetilde{\Phi}_2)u_R,
\ee
where $Q_L$ is the left-handed quark doublet, $u_R$ and $d_R$ are the right
handed quark singlets, $\widetilde{\Phi}_{1,2}= i \sigma_2 \Phi_{1,2}$ 
($\sigma_2$ is the Pauli matrix)
and $\Gamma_1^d$, $\Gamma_2^d$, $\Gamma_1^u$,
$\Gamma_2^u$
are matrices in the flavor space. One usually considers two versions
of the 2HDM:
\bea
\nonumber&&\Gamma_1^u=\Gamma_1^d=0 \hspace{1cm}({\rm Model~I});\\
\nonumber&&\Gamma_1^u=\Gamma_2^d=0 \hspace{1cm}({\rm Model~II}).
\eea
In this paper we also consider (neglecting FCNC's of the Lagrangian) 
the general case of the 2HDM (Model~III), where all the quantities 
$\Gamma_1^d$, $\Gamma_2^d$, $\Gamma_1^u$, $\Gamma_2^u$ are nonzero,
thus investigating possible impact of the Higgs doublets vacuum phase on the
$B \to X_d \gamma$ and $B \to X_d e^+ e^ - $ rate asymmetries.
Note also that the results obtained in
the Model~III in a large extent are valid for the multi-Higgs doublet models
with only one light charged Higgs particle \cite{3}. 

The physical charged 
Higgs state is related to the charged components of $\Phi_1$ and $\Phi_2$
as \cite{60}
\begin{displaymath}
\Phi_1^\pm = e^{ \pm i\delta} H^\pm \sin{\beta}, \hspace{0.5cm}
\Phi_2^\pm = H^\pm \cos{\beta}
\end{displaymath}
where 
$\tan{\beta}=v_2/v_1$ is the ratio of vev's of the Higgs doublets and
$\delta$ is the
Higgs doublets vacuum phase which yields spontaneous CP-violation.
The couplings of the charged Higgs boson
to $t_R$ and $b_R$ have the following form:
\begin{eqnarray}
\nonumber
\xi_t m_t H^ - (\bar{d},\bar{s},\bar{b})_L 
\left(\begin{array}{c} V^*_{td}\\ V^*_{ts}\\ V^*_{tb}\end{array}\right) 
t_R = 
ve^{i\delta_t}(h_{t_1}\sin{\beta} e^{ - i\delta} 
 -  h_{t_2} \cos{\beta})H^ - (\bar{d},\bar{s},\bar{b})_L 
\left(\begin{array}{c} V^*_{td}\\ V^*_{ts}\\ V^*_{tb}\end{array}\right) 
t_R  \\
\xi_b m_b H^+(\bar{u},\bar{c},\bar{t})_L 
\left(\begin{array}{c} V_{td}\\ V_{ts}\\ V_{tb}\end{array}\right) 
b_R = v e^{ - i\delta_b}(h_{b_1} \sin{\beta} e^{i\delta}  - 
h_{b_2}\cos{\beta})H^+(\bar{u},\bar{c},\bar{t})_L 
\left(\begin{array}{c} V_{td}\\ V_{ts}\\ V_{tb}\end{array}\right) 
b_R  
\end{eqnarray}
where $h_{t_{1,2}}=(\Gamma_{1,2}^u)_{33}$, 
$h_{b_{1,2}}= (\Gamma_{1,2}^d)_{33}$,
$v=\sqrt{v_1^2 + v_2^2}=174$GeV, V is the CKM matrix, $m_t$ and $m_b$
are the running top and bottom masses (for the
relations between the quarks running and pole masses see e.g. \cite{22})
and $\delta_t$, $\delta_b$  are related to
the parameters of the theory by \cite{11}
\begin{eqnarray*}
m_t e^{ - i \delta_t} = (h_{t_1} \cos{\beta}e^{ - i\delta} + 
h_{t_2} \sin{\beta}) v \\
m_b e^{i \delta_b} = (h_{b_1} \cos{\beta}e^{i\delta} + h_{b_2} \sin{\beta}) v
\end{eqnarray*}
We see that in the most general version of the 2HDM the couplings $\xi_t$ and $\xi_b$
are complex so that there is a new source of CP-violation connected with
spontaneous CP-breakdown. However in the Models~I~and~II these parameters
are real: $\xi_t=\xi_b= - \cot{\beta}$ and $\xi_t= - \cot{\beta}$, 
$\xi_b=\tan{\beta}$ respectively.

Some restrictions on
the couplings $h_{t_1}$, $h_{t_2}$, $h_{b_1}$, $h_{b_2}$ can be derived 
\cite{6} 
assuming the absence of new physics
\footnote{If the 2HDM originates as an "effective" low-energy
theory, then similar restrictions on the coupling constants of the theory can be
obtained, assuming the absence of new physics between the "full" theory
scale and unification scales.}
between the electroweak breaking scale $\mu_W\sim 100$GeV and 
unification scales $\mu_G \sim (10^{15} - 10^{19})$GeV and requiring the
validity
of the perturbation theory at the whole
energy range between $\mu_W$ and $\mu_G$.
As a result one obtains that at the electroweak breaking scale 
$h_{t_1}$, $h_{t_2}$, $h_{b_1}$, $h_{b_2}$ are of the order of unity or 
smaller. For the present experimental values of the top mass, 
$m_{t}^{pole}=(174 \pm 5)$GeV \cite{1}, it was found that $\tan{\beta} > 1$
for the Models~I~and~II.
It is easy to see from (2.1.2) that 
$|\xi_t| \sim 1$ or smaller. On the contrary one finds that
$|\xi_b|$ can be much larger than unity when $\tan{\beta} \gg 1$ or 
(in the Model~III) $\tan{\beta} \ll 1$. Strong
restrictions on $\xi_t$ and $\xi_b$ can also be derived from the experimental
constraints on ${\rm BR}(B\to X_s\gamma)$ and $B-\bar{B}$ mixing. We
discuss the experimental bounds on the parameters of the theory in subsection~2.3.

\subsection{Formulae for $B \to X_{d} \gamma$, $B \to X_{d} e^+ e^-$ decay
asymmetries.}
\renewcommand{\theequation}{2.2.\arabic{equation}}
\renewcommand{\thetable}{2.2.\arabic{table}}
\setcounter{equation}{0}

We study the decays $B \to X_{d(s)} \gamma$, $B \to X_{d(s)} e^+ e^-$  
using
the effective theory with five quarks obtained by integrating out
the heavy degrees of freedom which are
W and Z bosons, t-quark and the charged Higgs boson.
The effective Hamiltonian for  the decays $B \to X_{d(s)} \gamma$ can be
written as
\begin{eqnarray}
H_{eff}\Big(b \to d(s) \gamma(+g)\Big)=
-\frac{4G_F}{\sqrt{2}} \left [\lambda_{t}^{d(s)}
\sum_{i=1}^{8}C_i(\mu) O_i(\mu) 
-\lambda_u^{d(s)}\sum_{i=1}^{2}C_i(\mu) \Big(O_{i}^u(\mu)
- O_i(\mu)\Big) \right]    
\end{eqnarray}
where $\lambda_t^{d(s)}=V_{td(s)}^* V_{tb}$,
$\lambda_u^{d(s)}=V_{ud(s)}^* V_{ub}$. 
Here we present only the most important operators
\begin{eqnarray}
\nonumber
O_1 &=& \bar{d}(\bar{s})_L \gamma^\mu T^a c_L \ \bar{c}_L \gamma_\mu T^a b_L,
\hspace{1.3cm}
O_{1u} = \bar{d}(\bar{s})_L \gamma^\mu T^a u_L \ \bar{u}_L \gamma_\mu T^a b_L, \\
O_2 &=& \bar{d}(\bar{s})_L \gamma^\mu c_L \ \bar{c}_L \gamma_\mu b_L,
\hspace{2.25cm}
O_{2u} = \bar{d}(\bar{s})_L \gamma^\mu u_L \ \bar{u}_L \gamma_\mu b_L,  \\
\nonumber
O_7 &=& \frac{e}{16\pi^2} m_b(\mu) \ \bar{d}(\bar{s})_L
\sigma^{\mu \nu} b_R \ F_{\mu \nu}, \hspace{0.55cm}
O_8=\frac{g_s}{16\pi^2} m_b(\mu) \ \bar{d}(\bar{s})_L
\sigma^{\mu \nu} T^a b_R \ G_{\mu \nu}^a.
\end{eqnarray} 
The remaining operators contribute mainly through the 
mixing and otherwise can be safely neglected.
The full operator basis together with the expressions for
the corresponding  Wilson coefficients $C_i$ can be found
elsewhere \cite{14,3,13}.

The processes $\bar{B} \to \bar{X}_{d(s)} \gamma$
can be well described by the partonic level
transitions $b \to d(s) \gamma$ and $b \to d(s) \gamma g$. The
calculation of these partonic transitions includes the following three steps
\cite{16}:
\begin{enumerate}
\item{The Wilson coefficients $C_i$ at the heavy mass scale $\mu_W \sim M_W$
must be calculated, matching the effective and full theories. In the
next-to-leading order the matching has to be done at the $O(\alpha_s)$
level, i.e. $C_i(\mu_W)=C_i^{(0)}(\mu_W)+\alpha_s/(4\pi) C_i^{(1)}(\mu_W)$.}
\item{The renormalization group equations (RGE) must be used to obtain the
Wilson coefficients at the low-energy scale $\mu \sim m_b$. In the
next-to-leading order this step requires
the knowledge of the anomalous dimension matrix up to the order 
$\alpha_s^2$.}
\item{The matrix elements of the operators $O_i$ for the processes
$b \to d(s) \gamma$, $b \to d(s) \gamma g$ have to be calculated.}
\end{enumerate}

Since the effective Hamiltonian (2.2.1) is the same for the 2HDM and SM,
the 2HDM effects enter only through the first step.
At the $O(\alpha_s)$ level the matrix elements of $O_i$ for the processes
$b \to s \gamma$ and $b \to s\gamma g$
have been calculated in
ref's \cite{18,19,20,2}. The obtained results are relevant also for
$b \to d \gamma$ and $b \to d \gamma g$ decays when making an obvious
replacement $\lambda_i^s \to \lambda_i^d$, i=t,u.
At the order $\alpha_s^2$ the anomalous
dimension matrix can be found in \cite{14}. The explicit
relations between  the Wilson coefficients at low and high energy scales are
given in \cite{14,3}. Here are the leading order expressions
for the Wilson coefficients, corresponding
to the subset of the relevant operators (2.2.2)  
\footnote{More precisely,
these are the expressions for the effective Wilson
coefficients introduced in \cite{16}.}:
\begin{eqnarray}
\nonumber
C_1^{(0)}(\mu) & = & \tilde{\eta}^{6/23} - \tilde{\eta}^{-12/23}, \\
\nonumber
C_2^{'(0)}(\mu) &=&
C_2^{(0)}(\mu) - \frac{1}{6} C_1^{(0)}(\mu) =
\frac{1}{2} (\tilde{\eta}^{6/23} + \tilde{\eta}^{-12/23}), \\
C_7^{(0)}(\mu) &=&\tilde{\eta}^{16/23} C_7^{(0)}(\mu_W) + \frac{8}{3}
(\tilde{\eta}^{14/23}-\tilde{\eta}^{16/23}) C_8^{(0)}(\mu_W) +
\sum_{j=1}^{8} h_j \tilde{\eta}^{a_j} ,  \\ 
\nonumber
C_8^{(0)}(\mu) &=& \tilde{\eta}^{14/23} C_8^{(0)}(\mu_W) +
\sum_{j=1}^{8} h_j^{'} \tilde{\eta}^{a_j^{'}},
\end{eqnarray}
where $\tilde{\eta}=\alpha_s(\mu_W)/\alpha_s(\mu)$, and the numerical values of
$h_j$, $a_j$, $h_j^{'}$,
$a_j^{'}$ are available in \cite{14,3}. 

It is known that in the leading logarithmic
order the Wilson coefficients and
therefore ${\rm BR}(B \to X_{s(d)} \gamma)$ are highly sensitive to the
choice of the low-energy and matching scales mainly due to
dependence\footnote{$C_{7,8}^{(0)}(\mu)$ depend (via
$C_{7,8}^{(0)}(\mu_W)$) on the matching scale also due to dependence 
of the running top mass on the choice of $\mu_W$. However this dependence
leads only to about 3\% uncertainty in ${\rm BR}(B \to X_{s(d)} \gamma)$ \cite{21}.}
of $C_i^{(0)}(\mu)$ on  $\tilde{\eta}=\alpha_s(\mu_W)/\alpha_s(\mu)$.
However, in the next-to-leading order
the scale dependence of ${\rm BR}(B \to X_{s(d)} \gamma)$
in general is significantly reduced
\cite{16,21,2,14}.
 
At the matching scale
$\mu_W$ the coefficients  $C_i$, i=1,2,3,5,6, 
are the same as in the Standard
Model and can be found elsewhere \cite{21,12,3}. New Physics effects
contribute only to $C_4^{(1)}(\mu_W)$, $C_{7,8}^{(0)}(\mu_W)$ and 
$C_{7,8}^{(1)}(\mu_W)$. For $C_{7,8}^{(0)}(\mu_W)$ one has
\begin{eqnarray}
C_{7,8}^{(0)}(\mu_W) &=& C_{7,8}^{(0)^W} - \xi_t \xi_b C_{7,8}^{(0)^H} +
|\xi_t|^2 \tilde{C}_{7,8}^{(0)^H} 
\end{eqnarray}
where $C_{7,8}^{(0)^W}$, $C_{7,8}^{(0)^H}$, $\tilde{C}_{7,8}^{(0)^H}$
are given in ref. \cite{3}. The superscripts W, H denote respectively the
contribution from W-boson mediated and charged Higgs boson
mediated loops.

The direct CP-asymmetry in $B \to X_d \gamma$ decay is defined to be
\begin{equation}
a_{CP}(B \to X_d \gamma) = \frac{\Gamma(B \to X_d \gamma) -
\Gamma(\bar{B} \to \bar{X}_d \gamma)}{\Gamma(B \to X_d \gamma)+
\Gamma(\bar{B} \to \bar{X}_d \gamma)}.    
\end{equation}
It is known \cite{26}-\cite{5}, \cite{4}, \cite{10}
that the numerator of (2.2.5) starts at the order $O(\alpha_s)$. 
Since the order $O(\alpha_s^2)$ corrections to (2.2.5) are known
only partially, we expand (2.2.5) in $\alpha_s$ keeping only 
the leading term.
Consequently, we use one-loop RGE solution for $\alpha_s(\mu)$.

The method of extracting $a_{CP}(B \to X_{s(d)} \gamma)$ in the SM 
and its extensions is described in ref's \cite{26}-\cite{27}, \cite{10}. 
In the most general
form the expression for $B \to X_d \gamma$ decay asymmetry
is
(it is understood that the Wilson coefficients 
have to be taken at the low-energy scale $\mu$)
\begin{eqnarray}
\nonumber
 a_{CP}(B \to X_d \gamma)& = &\frac{2 \alpha_s(\mu)}
{|C_7^{(0)}|^2}\left\{\frac{2}{9}
{\rm Im}[C_7^{(0)*}C_8^{(0)}]+
\left(\frac{{\rm Re}[\lambda_u^d \lambda_t^{d {\large *}}]}
{|\lambda_t^d|^2}+1\right)
\frac{{\rm Im}[f_{27}]}{6\pi}C_2^{'(0)}{\rm Im}[C_8^{(0)}]- \right. \\
&&\hspace{-2.5cm}-\nonumber
  \left. \left[\left(\frac{20}{81}(1-r)+\frac{{\rm Im}[f_{27}]}{2\pi}\right)
\frac{{\rm Re}[\lambda_u^d \lambda_t^{d {\large *}}]}
{|\lambda_t^d|^2} - \frac{20}{81}r+\frac{{\rm Im}[f_{27}]}{2\pi} \right] 
C_2^{'(0)}{\rm Im}[C_7^{(0)}]+\right. \\ 
&&\hspace{-2.8cm}\nonumber
+\left. \frac{{\rm Im}[\lambda_u^d \lambda_t^{d {\large *}}]}
{|\lambda_t^d|^2}\left[\left(\frac{20}{81} (1-r)+
\frac{{\rm Im}[f_{27}]}{2\pi}\right)
C_2^{'(0)}{\rm Re}[C_7^{(0)}]- 
\frac{{\rm Im}[f_{27}]}{6\pi}C_2^{'(0)}{\rm Re}(C_8^{(0)})\right]-\right.\\
&&\hspace{1.4cm}\left.
-\frac{{\rm Im}[\lambda_u^d \lambda_t^{d {\large *}}]}
{|\lambda_t^d|^2}
\frac{{\rm Im}[f_{22}^{uc^*}]}{\pi}\left(C_2^{'(0)}\right)^2
\right\},
\end{eqnarray}
where r (the ratio of the absorptive parts of $<d\gamma|O_2 -
\frac{1}{6} O_1|b>$ and $<d\gamma|O_{2u} -
\frac{1}{6} O_{1u}|b>$ matrix elements) is given in \cite{10} and the
bremsstrahlung coefficients $f_{27}$ and $f_{22}$ can be taken from \cite{3}. 
The quantity $f_{22}^{uc^*}$ can be easily derived from the expression
for $f_{22}$ and reads 
\begin{displaymath}
f_{22}^{uc^*}=(f_{22}^{cu^*})^*=\frac{8 z}{27} \int_0^{1/z} dt (1 - zt)^2
\left(\frac{G^*(t)}{t} + \frac{1}{2}\right) \\
\end{displaymath}
where  $z=m_c^2/m_b^2$, and $G(t)$ can be found elsewhere \cite{18,20,2,3}.
Note that our formula for the $B \to X_{d(s)} \gamma$ rate asymmetry differs
from that in ref. \cite{27} by the last term in (2.2.6).
 This term arises from the interference of $<d\gamma
g|O_2 - \frac{1}{6} O_1|b>$ and $<d\gamma g|O_{2u} - \frac{1}{6}
O_{1u}|b>$ matrix elements.
 
The effective Hamiltonian for $B \to X_{d(s)} e^+ e^-$ decays
can be written as
\begin{equation}
H_{eff}(b \to d(s) e^+ e^-)=H_{eff}(b \to d(s) \gamma(+g)) -
\frac{4G_F}{\sqrt{2}}\lambda_{t}^{d(s)}(C_9 O_9 + C_{10} O_{10}) 
\end{equation}
where
\begin{eqnarray}
\nonumber
& & O_9=\bar{d}_L(\bar{s}_L)\gamma^{\mu} b_L \bar{e} \gamma_{\mu} e,
\hspace{0.5cm}
O_{10}=\bar{d}_L(\bar{s}_L)\gamma^{\mu} b_L \bar{e} \gamma_{\mu}
\gamma_5 e, \\
& & C_9^{(0)}(\mu) \approx P_0 + C_9^{(0)}(\mu_W), \hspace{0.5cm}
C_{10}^{(0)}(\mu)=C_{10}^{(0)}(\mu_W), \\   
\nonumber
& & C_{9,10}^{(0)}(\mu_W)=C_{9,10}^{(0)^W}(\mu_W) + 
|\xi_t|^2 C_{9,10}^{(0)^H}(\mu_W)
\end{eqnarray}
and $P_0$ and  $C_{9,10}^{(0)^{W,H}}(\mu_W)$ are given in \cite{15,52,54,49}.
We consider the dilepton invariant mass in the range
$1GeV^2 < s < 6GeV^2$: then the long-distance effects arising
from the contribution of
$\bar{c}c$ and $\bar{u}u$ resonances can be neglected so that
the decays
$\bar{B} \to \bar{X}_{d(s)} e^+ e^-$ are well described by
the transitions $b \to d(s) e^+ e^-$.
The (partially integrated) 
direct CP-asymmetry in the decay $B \to X_d e^+ e^-$ is defined by
\begin{equation}
a_{CP}(B \to X_d e^+ e^-) = \frac{\Delta B (B \to X_d e^+ e^-)
- \Delta B (\bar{B} \to \bar{X}_d e^+ e^-)}{\Delta B (B \to X_d e^+ e^-) +
\Delta B (\bar{B} \to \bar{X}_d e^+ e^-)} 
\end{equation}
where $\Delta B (\bar{B} \to \bar{X}_d e^+ e^-) =
\int_{1/m_b^2}^{6/m_b^2}{d\hat{s} \ dB(\bar{B} \to
\bar{X}_d e^+ e^-)/d\hat{s}}$,
$\hat{s} = s/m_b^2$. The quantity $dB(\bar{B} \to \bar{X}_d e^+ e^-)/d\hat{s}$
can be expressed in terms of $C_7^{(0)}(\mu)$ and
$C_9^{eff}(\mu) = C_9^{(0)}(\mu)(1 + \omega(\hat{s})) + Y_d(\hat{s})$,
where $\omega(\hat{s})$ and $Y_q(\hat{s})$,
q=s,d, are given in ref. \cite{54}. Denoting 
$C_9^{eff}(\mu) = \xi_1 +
[(\lambda_u^d \lambda_t^{d*})/|\lambda_d^t|^2] \xi_2$ one gets
for $B \to X_d e^+ e^-$ decay rate asymmetry
\begin{eqnarray}
\nonumber
a_{CP}&=&\int d\hat{s} (\hat{s} - 1)^2
\left\{2 \frac{{\rm Im}[\lambda_u^d\lambda_{t}^{d*}]}{|\lambda_t|^2}
\Big[(1+2 \hat{s}){\rm Im}[\xi_1^* \xi_2] + 
6 {\rm Re}[C_7^{(0)}]{\rm Im}[\xi_2]\Big] \right.\\
&&\left.\hspace{1.8cm}- 12 \frac{{\rm Re}(\lambda_u^d\lambda_{t}^{d*}]}
{|\lambda_t|^2}
{\rm Im}[C_7^{(0)}]{\rm Im}[\xi_2] -
12 {\rm Im}[C_7^{(0)}]{\rm Im}[\xi_1] \right\}{\Big/}  \\      
\nonumber
&&\hspace{-1.cm}{\Big/}\left\{\int d\hat{s}
(\hat{s} - 1)^2\Big((1+2\hat{s})\Big[|C_9^{eff}|^2 +
|C_{10}|^2\Big]  
\nonumber
+ 12 {\rm Re}[C_7^{(0)} \ C_9^{eff*}] +
4 \frac{2+\hat{s}}{\hat{s}} |C_7^{(0)}|^2
\Big)\right\},
\end{eqnarray}
where the limits of the integrations are specified above.

\subsection{Constraints on parameters of theory.}
\renewcommand{\theequation}{2.3.\arabic{equation}}
\renewcommand{\thetable}{2.3.1\arabic{table}}
\setcounter{equation}{0}
We use the following values of the parameters of the theory: $M_W=80$GeV,
$m_t^{pole}=(174 \pm 5)$GeV, $\alpha_s(M_Z)=0.119 \pm 0.002$ \cite{1},
$m_b^{pole}=(4.8 \pm 0.15)$GeV, $m_c/m_b=0.29 \pm 0.02$ (corresponding to
$r=0.145_{-0.03}^{+0.035}$, ${\rm Im}[f_{27}]/(2\pi)=0.011 \pm 0.001$,
${\rm Im}[f_{22}^{uc^*}]/\pi=0.0055 \pm 0.0013$),
$\alpha_{em}=1/(130.3 \pm 2.3)$ and $BR_{sl}=
{\rm BR}(B \to X_c e^+ \nu)=(10.49 \pm 0.46)\%$ \cite{2,3,4}.
The low-energy scale $\mu$
and the matching scale $\mu_W$ are
varied between $m_b/2$, $2m_b$ and $M_W$, $m_t$ respectively.

In the Wolfenstein parameterization  the necessary CKM-factors
$\lambda_u^{d(s)}$,$\lambda_t^{d(s)}$ and $V_{cb}$  are given by
\cite{4,5}:
\begin{eqnarray}
\nonumber
\lambda_t^d &=& A \lambda^3 (1 - \bar{\rho} + i \bar{\eta}), \hspace{2.84cm}
\lambda_u^d=A \lambda^3 (\bar{\rho} - i \bar{\eta}) ,
\hspace{0.5cm}V_{cb}=A \lambda^2,\\
\lambda_t^s &=& -A \lambda^2 (1- \lambda^2/2 + \lambda^2 (\rho -
i \eta)),  
\hspace{0.5cm}\lambda_u^s = A \lambda^4 (\rho - i \eta),
\end{eqnarray}
where $\bar{\rho}=(1- \frac{\lambda^2}{2}) \rho$,
$\bar{\eta}=(1- \frac{\lambda^2}{2}) \eta$,
$\lambda=\sin{\theta_C} \approx 0.22$ and A=$0.819 \pm 0.035$.
The restrictions on the parameters $\rho$ and $\eta$ can be obtained from
the unitarity fits, which yield \cite{7,8}
\begin{eqnarray}
\sqrt{\rho^2 + \eta^2} = 0.423 \pm 0.064, \hspace{0.5cm}
38^o \leq {\bar\gamma} \leq 81^o \\ 
16^o \leq {\bar\beta} \leq 34^o, \hspace{0.5cm}
75^o \leq {\bar\alpha} \leq 121^o 
\end{eqnarray}
where ${\bar\alpha}$, ${\bar\beta}$ and ${\bar\gamma}$ are the
angles of the unitarity triangle.
The conditions (2.3.2) lead to the following constraints on $\rho$ and
$\eta$:
\begin{equation}
0.05 \leq \rho \leq 0.38, \hspace{0.5cm}
0.22 \leq \eta \leq 0.48                  
\end{equation}
During the numerical analysis we use (2.3.2) and (2.3.3) rather than (2.3.4)
since it turns out that the correlations between the values of $\rho$ and
$\eta$ reduce significantly the derived range for the 
CP-asymmetry\footnote{Note that the  2HDM fits for the parameters $\rho$ and
$\eta$ can in general be different from those of the SM.  
This source of  deviations from the 
SM predictions however is out of the scope of the present paper.}. 

The restrictions on the parameters $\xi_t$ and $\xi_b$ coming from the bounds
on t- and b-quark Yukawa couplings have been discussed in subsection~2.1.
It has been pointed out that while $|\xi_t| \sim 1$ or smaller, $|\xi_b|$
can be much larger than unity when $\tan{\beta} \gg 1$ or 
$\tan{\beta} \ll 1$.
This means that if no other constraints on the 2HDM
parameters are imposed, the Wilson coefficients 
$C_7^{(0)}(\mu)$ and $C_8^{(0)}(\mu)$ can be significantly larger than
in the Standard Model.
The additional constraints on $C_7^{(0)}(\mu)$ and
$C_8^{(0)}(\mu)$ can be obtained from the experimental bound on
$B \to X_s \gamma$ branching.
We require that
\begin{equation}
{\rm BR}^{2HDM}(<B \to X_s \gamma>) = {\rm BR}^{exp}(<B \to X_s \gamma>) =
(3.15 \pm 0.54) \times 10^{-4} 
\end{equation}
where ${\rm BR}^{2HDM}(<B$$\to X_s\gamma>)$ is calculated in the next-to-leading
order according the formulae given in ref. \cite{3}. Due to NLO effects
the restrictions on $|C_7^{(0)}(\mu)|$ coming from (2.3.5) can differ sizably
from those given in \cite{10}: for instance, in the case when 
${\rm sign}({\rm Re}[C_7^{(0)}(\mu)])= -{\rm sign}(C_7^{(0)SM}(\mu))$, 
$|C_7^{(0)}(\mu)|$ can be up to 3 times larger than in the SM.
Some deviation of $|C_7^{(0)}(\mu)|$ from $|C_7^{(0)SM}(\mu)|$ can
also occur due to the fact that the SM prediction for the $B \to X_s \gamma$
branching ratio only overlaps with the experimental data.

As one can deduce from (2.2.4) and the relation 
$|\tilde{C}_{7}^{(0)H}| \ll |C_{7}^{(0)H}|$ \cite{3,11}, the condition 
(2.3.5)
restricts the product $\xi_t \xi_b$, whereas the parameter $\xi_t$ and
hence the Wilson coefficients $C_9^{(0)}(\mu)$ and 
$C_{10}^{(0)}(\mu)$ are still constrained only due to the bounds on t- and
b- quark Yukawa couplings. Additional restrictions on
$|\xi_t|$  (respectively on $C_{9,10}^{(0)}(\mu)$) can be
obtained from the experimental  measurements of the $B-\bar{B}$ mixing.
For the central values of the parameters entering the expression
for $\Delta m_B$ one obtains that $|\xi_t| \leq 0.3$ when
$m_{H^+} \sim 100$GeV \cite{56}. However when taking into account
experimental
uncertainties of these parameters (especially important are those of
$f_B \sqrt{B_B}= (215 \pm 40)$MeV \cite{7}) one derives much weaker
constraint.
Thus, for   $m_{H^+} = 100$GeV, $\rho = 0.19$ and  $\eta=0.37$ we
obtain that $|\xi_t| \leq 0.9$. The restriction on $|\xi_t|$ again
becomes stronger when $\rho \to 0$: $|\xi_t| \leq 0.7$ now. On the contrary,
for  $\rho \to \rho_{max}=0.38$ the bound on $|\xi_t|$ coming from the
experimental  constraints on $B-\bar{B}$ mixing becomes inessential.

The restrictions on the parameters $\xi_t$ and $\xi_b$ are also correlated
with the charged Higgs mass. In our calculations we vary it as 
$100GeV \leq m_{H^+} \leq 400GeV$. For larger values 
of $m_{H^+}$ the theoretical predictions for ${\rm BR}(B \to X_s \gamma)$
(and hence the restrictions on $C_7^{(0)}(\mu)$ and
$C_8^{(0)}(\mu)$) become inaccurate, as the formulae for
${\rm BR}(B \to X_s \gamma)$ contain terms proportional to 
$\log{\frac{\mu_W^2}{m_{H^+}^2}}$ \cite{3}. 

Some bounds on the 2HDM parameters could be obtained also from the
experimental constraints on the neutron electric dipole
moment (NEDM).
Note however that these constraints are known to suffer from
the uncertainty of the hadronic matrix element $<O_g(\mu_n)>$ 
\cite{23}, where $O_g=\frac{1}{6} f^{abc}
\epsilon^{\delta \nu \alpha \beta} G^a_{\alpha \beta} G^b_{\lambda \delta}
G^c_{\lambda \nu}$ (here G denotes the gluon field strength tensor).
When taking into account the uncertainty in the choice of $<O_g(\mu_n)>$
(unlike \cite{56} where a particular value of $<O_g(\mu_n)>$ is used)
the NEDM constraints restrict ${\rm Im}[\xi_t \xi_b]$ 
weaker than the condition (2.3.5).


\section{Direct CP-asymmetry in the  decay $B \to X_d \gamma $.}
\setcounter{equation}{0}
\setcounter{table}{0}
\renewcommand{\theequation}{3.\arabic{equation}}
\renewcommand{\thetable}{3.\arabic{table}}

Though the condition (2.3.5) strongly restricts the 2HDM parameter
space, the predictions of this model for the
$B \to X_d \gamma$ decay asymmetry can deviate significantly from the
result derived in the Standard Model. This
is mainly due to the fact that the condition (2.3.5) puts
constraints on $|C_7^{(0)}(\mu)|$ while its phase
remains in general free. In particular, even being real
$C_7^{(0)}(\mu)$ and consequently $a_{CP}(B \to X_d \gamma)$
can have the sign opposite to that in the SM. 
In other words the measurement of the direct CP-asymmetry in
$B \to X_d \gamma$ decay offers an alternative to the method
proposed in \cite{24,55} where the authors suggest to use  
$B \to X_s e^+ e^-$ branching and 
forward - backward asymmetry to determine the sign of
$C_7^{(0)}(\mu)$.
The deviations
of the 2HDM predictions from those of the SM can occur also
due to the contribution of the Higgs doublets vacuum phase which
gives rise the terms of (2.2.6) proportional to
${\rm Im}[C_7^{(0)}(\mu)]$ and ${\rm Im}[C_8^{(0)}(\mu)]$. This contribution 
can lead both to the change of sign of $a_{CP}$ and to the
increasing or suppression of its absolute value as compared to that in the
SM. So in the two-Higgs doublet extension of the Standard Model there are
two sources of deviations of the predictions for $a_{CP}(B \to X_d \gamma)$
from those of the SM. These sources are connected respectively
with the change of sign of ${\rm Re}[C_7^{(0)}(\mu)]$, as compared to that in
the SM, and with
the presence of an additional (to the CKM one) weak phase in theory.
\begin{table}
\caption{$a_{CP}(B \to X_d \gamma)$ in the Model I.}
\begin{center}
\begin{tabular}{|c|c|c|c|c|c|c|}
\hline
 & \multicolumn{6}{c|}{$a_{CP}(B \to X_d \gamma)/10^{-2}$} \\
\cline{2-7} $m_{H^+}$, GeV & \multicolumn{3}{c|}{$\mu_W=M_W$} &
\multicolumn{3}{c|}{$\mu_W=m_t$} \\
\cline{2-7} & $\mu=m_b/2$ & $\mu=m_b$ & $\mu=2 m_b$
& $\mu=m_b/2$ & $\mu=m_b$ & $\mu= 2 m_b$ \\
\hline
100 & $13 \div 36$ & $11 \div 33$ & $10 \div 30$ &
$13 \div 32$ & $11 \div 30$ & $10 \div 27$  \\  \hline
200 & $13 \div 35$ & $11 \div 33$ & $10 \div 30$ &
$13 \div 32$ & $11 \div 30$ & $10 \div 27$  \\  \hline
400 & $13 \div 31$ & $11 \div 28$ & $10 \div 26$ &
$13 \div 31$ & $11 \div 28$ & $10 \div 26$  \\  \hline
\end{tabular}
\end{center}
\end{table}
In the Model~III both of sources of deviations from the SM predictions are
actual hence we investigate their interplay in details. On the contrary
in the Models~I~and~II both of these sources are irrelevant. Indeed, 
as it was discussed in the previous section, 
the parameters $\xi_t$, $\xi_b$ and hence $C_7^{(0)}(\mu)$ and
$C_8^{(0)}(\mu)$ are real. Furthermore, for $\tan{\beta} > 1$ the sign
of $C_7^{(0)}(\mu)$ is the same as in the SM (i.e.
$C_7^{(0)}(\mu) < 0$). As a consequence in the Models~I~and~II the
predictions for $a_{CP}(B \to X_d\gamma)$ 
are close to those of the Standard Model.

Varying the parameters of the theory within
the intervals specified in subsection~2.3 one finds that in the SM
$a_{CP} (B \to X_d \gamma) = (10 \div 27)\%$. The uncertainty in the
determination of $a_{CP}$ is connected mainly with rather large dispersion of
values of CKM parameters $\rho$ and $\eta$ (for the detailed analysis of the
behavior of $a_{CP} (B \to X_d \gamma)$ with $\rho$ and $\eta$ see ref.
\cite{4}). Indeed, if we fix these
parameters e.g. at their "best fit" values, $\rho=0.19$, $\eta=0.37$,
then the uncertainty for the decay asymmetry is
reduced significantly: one gets 
$a_{CP} (B \to X_d \gamma) = (14 \div 20)\%$  now.
The decay asymmetry
shows moderate dependence ($\sim 15\%$) on the low-energy scale $\mu$.
Surprisingly the dependence of $a_{CP} (B \to X_d \gamma)$ on the 
matching scale $\mu_W$ is negligible.

Note that obtained maximum (minimum) value of  $B \to X_d \gamma$ decay
asymmetry is about 1.3 (1.5) times smaller (larger) than the one reported
in ref. \cite{4}. As it is mentioned above, this is mainly
due to taking into
account the correlations between the restrictions on $\rho$ and $\eta$.
\begin{table}
\caption{$a_{CP}(B \to X_d \gamma)$ in the Model III.}
\begin{center}
\begin{tabular}{|c|c|c|c|c|c|c|}
\hline
 & \multicolumn{6}{c|}{$a_{CP}(B \to X_d \gamma)/10^{-2}$} \\
\cline{2-7} $m_{H^+}$, GeV & \multicolumn{3}{c|}{$\mu_W=M_W$} &
\multicolumn{3}{c|}{$\mu_W=m_t$} \\
\cline{2-7} & $\mu=m_b/2$ & $\mu=m_b$ & $\mu=2 m_b$
& $\mu=m_b/2$ & $\mu=m_b$ & $\mu= 2 m_b$ \\
\hline
100 & $-19 \div 37$ & $-10 \div 34$ & $-7 \div 30$ &
$-25 \div 33$ & $-15 \div 31$ & $-10 \div 27$  \\ \hline
200 & $-17 \div 37$ & $-10 \div 34$ & $-7 \div 30$ &
$-25 \div 34$ & $-15 \div 31$ & $-10 \div 27$  \\ \hline
400 & $-16 \div 37$ & $-9 \div 34$ & $-7 \div 30$ &
$-24 \div 34$ & $-15 \div 32$ & $-10 \div 27$  \\ \hline
\end{tabular}
\end{center}
\end{table}
\begin{table}
\caption{$a_{CP}(B \to X_d \gamma)$ in the Model~III for
$m_{H^+}=100$GeV, $\xi_t=0.5$ and a) $\xi_t \xi_b=4$ - the
case when $C_7^{(0)}(\mu)$ is real and positive;
 b) $\xi_t \xi_b= 2 \pm 2i$ -
the simplest case when $arg(C_7^{(0)}(\mu)) \to \pm \pi/2$.}
\begin{center}
\begin{tabular}{|c|c|c|c|c|c|c|c|}
\hline
  &  & \multicolumn{3}{c|}{$\mu_W=M_W$} &
\multicolumn{3}{c|}{$\mu_W=m_t$} \\
\cline{3-8} & &$\mu=m_b/2$ & $\mu=m_b$ & $\mu=2 m_b$
& $\mu=m_b/2$ & $\mu=m_b$ & $\mu= 2 m_b$ \\
\hline
$a_{CP}/10^{-2}$ & a) &
 $-23 \div -10$ & $- 14\div -6$ & $-9 \div -5$ &
$-28 \div -13$ & $-16 \div -8$ & $-10 \div -5$  \\ 
\cline{2-8} & b) &
 $-10 \div 14$ & $- 10\div 6$ & $-8 \div 2.5$ &
$-10 \div -18$ & $-10 \div 8$ & $-9 \div 4$  \\ \hline
\end{tabular}
\end{center}
\end{table}

As it is noted above, in the Models~I~and~II the
contribution of the New Physics  to the decay asymmetry is not sizable. 
In the Model~I $a_{CP}(B \to X_d \gamma)=(10 \div 36)\%$ as one
can see from Table~3.1. In the Model~II one must take 
$m_{H^+} > 400$GeV when applying (2.3.5): then, as the numerical 
analysis shows, $a_{CP}(B \to X_d \gamma)$ remains within the 
SM interval.
 
In the Model~III the predictions
for the $B \to X_d \gamma$ decay asymmetry can differ from those
of the SM significantly.
As one can see from Table~3.2, $a_{CP} (B \to X_d \gamma)$ can be negative
now. Furthermore it can be arbitrary small in absolute value
and hence unmeasurable.  Large deviations
from the SM predictions concern principally the minimum value of
$B \to X_d \gamma$ decay asymmetry.
As for the maximum value of CP-asymmetry, 
it can be larger than in the Standard Model only about $1.3$
times. The deviation of
$a_{CP}^{max} (B \to X_d \gamma)$ from the SM prediction is due to
the fact that the experimentally allowed range for ${\rm BR}(B \to X_s \gamma)$
only overlaps with the SM predictions.

Thus we obtain that in the Model~III 
$a_{CP}(B \to X_d \gamma)= (-25 \div 37)\%$. 
Note that the maximum (absolute) value of CP-asymmetry is 
1.5 - 2 times larger than in \cite{10}. This is due to the use of new
experimental results for the CKM parameters, taking into 
account the gluon bremsstrahlung effects and the scale uncertainty of
the obtained results as well as due to the fact that 
our constraints on $C_7^{(0)}(\mu)$ differ from those of
\cite{10}.

We investigate in details the scale dependence of  
$a_{CP}(B\to X_d\gamma)$. As one can see from Table~3.2,
minimum value of $a_{CP}(B\to X_d\gamma)$ is highly sensitive to
the choice of low-energy and matching scales: it varies from $-25\%$
to $-7\%$ when varying $\mu$ and $\mu_W$ between $m_b/2$, $2 m_b$
and  $m_t$,$M_W$ respectively.
In Table 3.3 we present the derived
interval of $B\to X_d\gamma$ decay rate 
asymmetry for the fixed values of the parameters $\xi_t$ and $\xi_b$.
It is easy to see that large $\mu$ and $\mu_W$ dependence 
for $a_{CP}(B\to X_d\gamma)$ is observed both when
$C_7^{(0)}(\mu)$ changes the sign, as compared to the SM, and
when the effect of Higgs doublet vacuum phase becomes sizable.
So, unlike the SM,
in the 2HDM the direct CP-asymmetry in $B \to X_d \gamma$
decay can be highly sensitive to the choice of the renormalization scale
when the New Physics contribution is sizable.

\begin{figure}[t]
\begin{flushleft}
\leavevmode
\epsfxsize=7.5cm
\epsffile{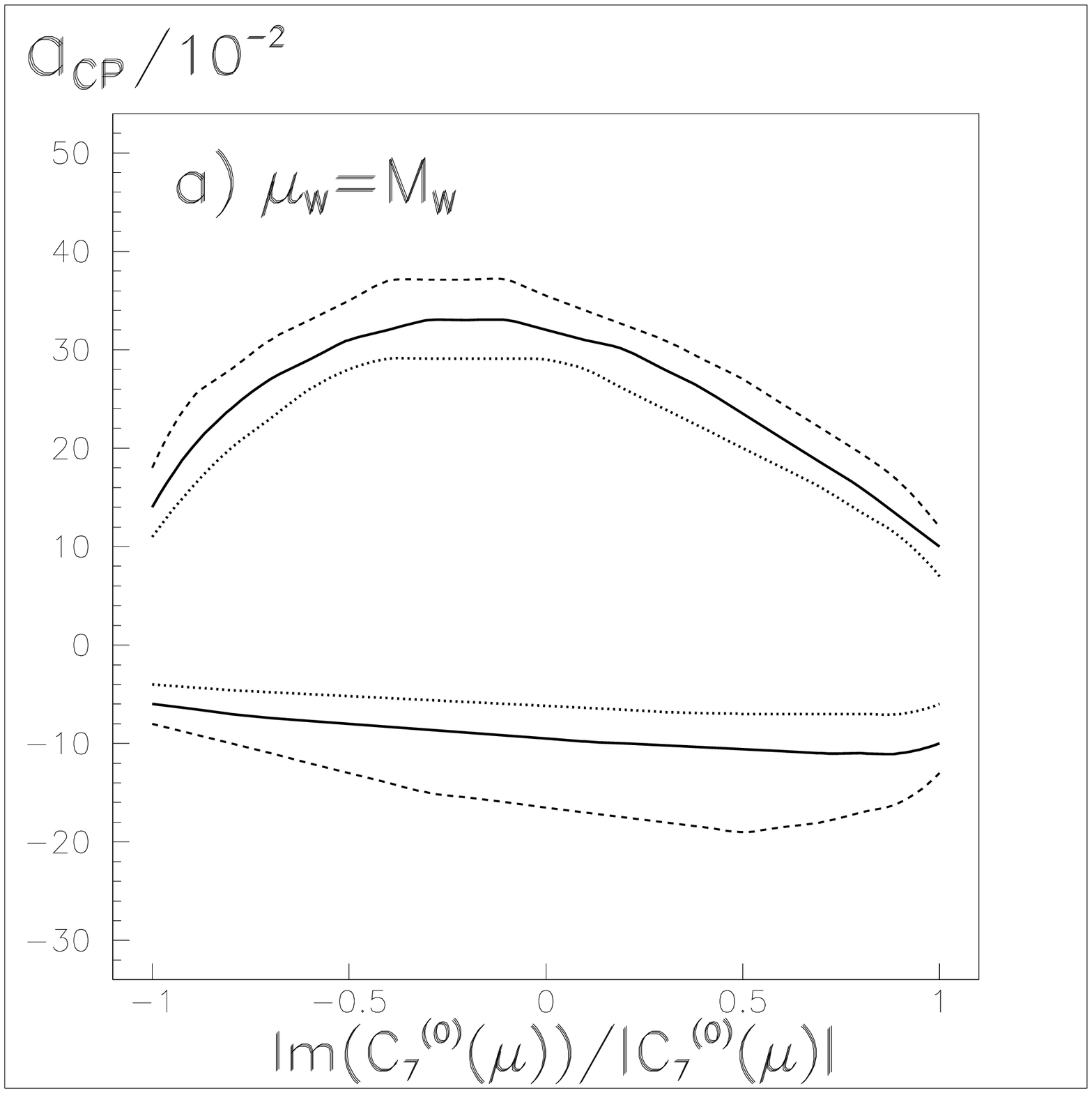}
\end{flushleft}
\vspace{-11.33cm}
\begin{flushright}
\leavevmode
\epsfxsize=7.5cm
\epsffile{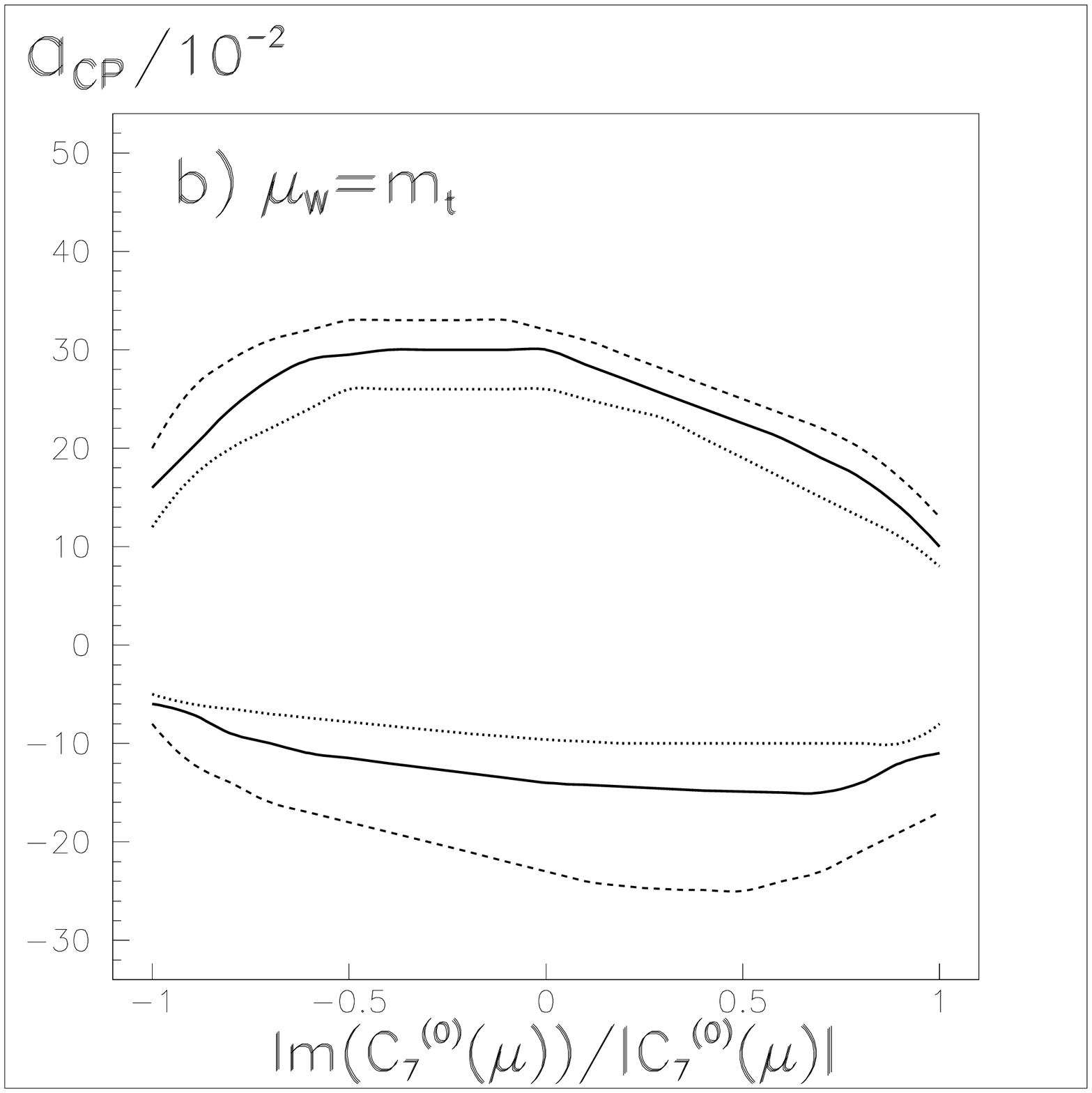}
\end{flushright}
\vspace{-1.5cm}
\caption{Maximum and minimum values of $a_{CP}(B \to X_d \gamma)$ as
functions of ${\rm Im}(C_7^{(0)}(\mu))/C_7^{(0)}(\mu)|$ 
for $m_{H^+}=100$GeV and
$\mu=m_b/2$ (dashed line), $\mu=m_b$ (solid line), $\mu=2 m_b$ 
(dotted line).}
\end{figure}

Finally we examine the influence of the Higgs doublet vacuum phase
on the obtained results. 
For this purpose we present in  Fig.~1
$a_{CP}(B \to X_d \gamma)$ as a function of 
${\rm Im}(C_7^{(0)}(\mu)/|C_7^{(0)}(\mu)|$. Though the obtained results
are renormalization scale dependent, they have the following common
feature. As one can see from Fig.~1, maximum (minimum) value of 
$a_{CP}(B \to X_d \gamma)$ derived for\footnote{It is easy to see
from (2.2.4) that in this case ${\rm Im}(C_8^{(0)}(\mu))=0$ as well.}
${\rm Im}(C_7^{(0)}(\mu))=0$
is close or coincides with the absolute maximum (minimum).
On the contrary, for ${\rm Im}(C_7^{(0)}(\mu)/|C_7^{(0)}(\mu)|\approx 1$ 
the allowed interval of CP-asymmetry shrinks about 1.5 times: 
$a_{CP}(B \to X_d \gamma)=(-17 \div 20)\%$ now. Thus
one may conclude that in the 2HDM the new source of CP-violation
either holds the allowed interval of $B \to X_d \gamma$ decay asymmetry
unaltered or leads to the suppression of $|a_{CP}(B \to X_d \gamma)|$.

It is important to stress that because of the Higgs doublet vacuum
phase effects the $B \to X_d \gamma$ decay asymmetry can be arbitrary
small in absolute value and hence unmeasurable. 
This is in contrast to the case with real
$C_7^{(0)}(\mu)$ and $C_8^{(0)}(\mu)$ when there is a lower bound on the
$B \to X_d \gamma$ decay asymmetry: then
$a_{CP}(B \to X_d \gamma) =(-24 \div -3)\%$ when being negative,
and $a_{CP}(B \to X_d \gamma) =(10 \div 36)\%$ when being positive.
Notice  also
that the behavior of $a_{CP}(B \to X_d \gamma)$ with 
${\rm Im}[(C_7^{(0)}(\mu)]$ is different from that of 
$a_{CP}(B \to X_s \gamma)$. As it was shown 
in ref's. \cite{11,10,3}, due to the presence of new source of
CP-violation connected with spontaneous CP-breakdown
the absolute value of $a_{CP}(B \to X_s \gamma)$ 
in the 2HDM can be a few times
larger than in the SM. 
 
One can also expect that the behavior of $a_{CP}(B \to X_d \gamma)$
with respect to any new source of CP-violation will be
similar to that in the Model~III for any extension of
the Standard Model where the New Physics contribution to $C_8^{(0)}(\mu)$ is
of the order or smaller than that to $C_7^{(0)}(\mu)$. Notice however
that in the models with enhanced chromomagnetic moment operator
the effect of new phases can be sizable \cite{27}.  Particular
examples of such models are the supersymmetric theories. 
So far detailed analysis has been performed only for 
$B \to X_s \gamma$ rate asymmetry \cite{45}-\cite{42}.
The investigation
of $B \to X_d \gamma$ decay asymmetry in the MSSM has been carried out 
only for the case when the supersymmetry
breaking parameters are real \cite{5}.
It is interesting also to examine the case
when these parameters are complex (and their phases are not suppressed).
This work is in progress now.

\section{Direct CP-asymmetry in the decay $B \to X_d e^+ e^-$.}
\setcounter{equation}{0}
\setcounter{table}{0}
\renewcommand{\theequation}{4.\arabic{equation}}
\renewcommand{\thetable}{4.\arabic{table}}

The deviations of $a_{CP}(B \to X_d e^+ e^-)$ from the SM
predictions can occur due to New Physics contribution to 
$C_7^{(0)}(\mu)$, $C_9^{(0)}(\mu)$, $C_{10}^{(0)}(\mu)$.
One must however take into account that the 2HDM effects modify these
Wilson coefficients in different ways.
One may use for $C_9^{(0)}(\mu)$ the SM
prediction since
numerically $C_9^{H}(\mu) < 0.025C_9^{W}(\mu)$. On the contrary the
deviation of $C_{10}^{(0)}(\mu)$ from its SM value can be sizable.
The Higgs-mediated loops contribution makes
$|C_{10}^{(0)}(\mu)|$ and hence the denominator of (2.2.10) larger than in
the SM. Note however that the deviations of $C_{10}^{(0)}(\mu)$ from
$C_{10}^{(0){SM}}(\mu)$ depend on the restrictions on the parameter
$\xi_t$ and hence on the choice of the CKM parameters. These
deviations are the largest for $\rho \to \rho_{max}$ and negligible for
$\rho \to \rho_{min}$. As the numerical analysis
shows, in the Standard Model the latter choice of $\rho$ corresponds to
the minimum value of $B \to X_d e^+ e^-$ rate asymmetry. As a result, 
the New Physics contribution to
$C_{10}^{(0)}(\mu)$ holds the allowed interval of the
$a_{CP}(B \to X_d e^+ e^-)$ unaltered as compared to that in 
the SM (note that with the definition (2.2.10) $a_{CP}(B \to X_d e^+ e^-)$
is positive in the SM).

It is easy to see that if 
$|C_7^{(0)}(\mu)| \sim |C_7^{(0)^{SM}}(\mu)| \ll C_9^{(0)}(\mu)$,
the New Physics contribution to $C_7^{(0)}(\mu)$ has negligible impact
on $B \to X_d e^+ e^-$ rate asymmetry. In the case when 
$|C_7^{(0)}(\mu)| > |C_7^{(0)^{SM}}(\mu)|$ (generally speaking in this
case ${\rm Re}(C_7^{(0)}(\mu)) > 0$) the denominator of (2.2.10) 
again increases. The numerator of (2.2.10) can be
both larger and smaller than in the SM.
Possible enhancement of the numerator
cancels with the enhancement of the denominator - thus in the 2HDM the direct 
CP-asymmetry in $B \to X_d e^+ e^-$ decay in general does not
exceed by more than $(10 \div 15)\%$ its maximum value in the SM.
However $a_{CP}(B \to X_d e^+ e^-)$ can be significantly smaller than
in the SM. Thus we disagree with the result of ref. \cite{49},
according to which the effects of New Physics  in general make
$a_{CP}(B \to X_d e^+ e^-)$ larger than in the SM.

\begin{table}
\caption{$a_{CP}(B \to X_d e^+ e^-)$ in the SM for
different choices of $\mu$ and $\mu_W$.}
\begin{center}
\begin{tabular}{|c|c|c|c|c|c|c|}
\hline
 & \multicolumn{3}{c|}{$\mu_W=M_W$} &
\multicolumn{3}{c|}{$\mu_W=m_t$} \\
\cline{2-7} & $\mu=m_b/2$ & $\mu=m_b$ & $\mu=2 m_b$
& $\mu=m_b/2$ & $\mu=m_b$ & $\mu= 2 m_b$ \\
\hline
$a_{CP}/10^{-2}$ &  $1 \div 2.4$ & $3 \div 6.4$ & $4.9 \div 9.8$ &
$0.2 \div 0.8$ & $2.3 \div 5$ & $4.2 \div 8.7$ \\
\hline
\end{tabular}
\end{center}
\end{table}
\begin{table}
\caption{$a_{CP}(B \to X_d e^+ e^-)$ in the Model~III for 
$m_{H^+}=(100 \div 400)$GeV and
different choices of $\mu$ and $\mu_W$.}
\begin{center}
\begin{tabular}{|c|c|c|c|c|c|c|}
\hline  & \multicolumn{3}{c|}{$\mu_W=M_W$} &
\multicolumn{3}{c|}{$\mu_W=m_t$} \\
\cline{2-7} & $\mu=m_b/2$ & $\mu=m_b$ & $\mu=2 m_b$
& $\mu=m_b/2$ & $\mu=m_b$ & $\mu= 2 m_b$ \\
\hline
$a_{CP}/10^{-2}$ & $0.4 \div 2.7$ & $1.9 \div 6.9$ & $2.6 \div 10.7$ &
$-0.4 \div 1.3$ & $1.4 \div 5.4$ & $2.6 \div 9.3$  \\ \hline
\end{tabular}
\end{center}
\end{table}
Let us proceed to concrete results. The derived interval of
$a_{CP}(B \to X_d e^+ e^-)$ in the Standard Model
for different choices of the low-energy
and matching scales is given in Table~4.1. As one can see, the 
direct CP-asymmetry in the decay $B \to X_d e^+ e^-$ is highly
sensitive to the choice of the low energy scale: one gets $\sim 65\%$
uncertainty in the obtained results when varying $\mu$ from $m_b/2$ to
$2 m_b$. This uncertainty can be much larger and exceed 100\% when
examining the renormalization scale dependence of
$a_{CP}(B \to X_d e^+ e^-)$ with the use of so-called Kagan-Neubert
 approach \cite{48,62}.
It is interesting that varying the matching scale from $M_W$
to $m_t$, one obtains that the predictions for CP-asymmetry are shifted
by the same magnitude independently of the choice of $\mu$:  by about
$(-0.7) \times 10^{-2}$ for the minimum value and
$-(1.2 \div 1.5) \times 10^{-2}$ for the 
maximum value.

For the fixed values of the low-energy and matching scales 
the uncertainty of
$a_{CP}(B \to X_d e^+ e^-)$ in the SM is connected mainly with
large dispersion of the CKM parameters $\rho$ and $\eta$, analogously to
$a_{CP}(B \to X_d \gamma)$. For the "best fit" values 
$\rho=0.19$, $\eta=0.37$, $a_{CP}(B \to X_d e^+ e^-)=(4.3 \pm 0.4)\%$. 
The behavior of $B \to X_d e^+ e^-$ decay
asymmetry with $\rho$ (equivalently, $\eta$) for different choices of
$\sqrt{\rho^2 + \eta^2}$ and for $\mu=m_b$, $\mu_W = M_W$ is presented in
Fig.~2. As one can see from this figure, $a_{CP}(B \to X_d e^+ e^-)$ 
increases as $\rho$ or $\sqrt{\rho^2 + \eta^2}$ increases.

The renormalization scale dependence makes the  results for 
$a_{CP}(B \to X_d e^+ e^-)$ in the SM ambiguous:
when choosing $\mu = 2 m_b$ and $\mu_W = M_W$, the direct
CP-asymmetry in $B \to X_d e^+ e^-$ decay can reach 10\%. On the other
hand, for $\mu = m_b/2$ and $\mu_W = m_t$, $a_{CP}(B \to X_d e^+ e^-)$
is smaller than 1\% thus being hardly measurable in future experiments.
This ambiguity preserves also in the two-Higgs doublet extension of the SM.

\begin{figure}[t]
\begin{center}
\epsfysize=15cm
\epsfysize=10cm
\vspace{-1cm}
\epsfbox{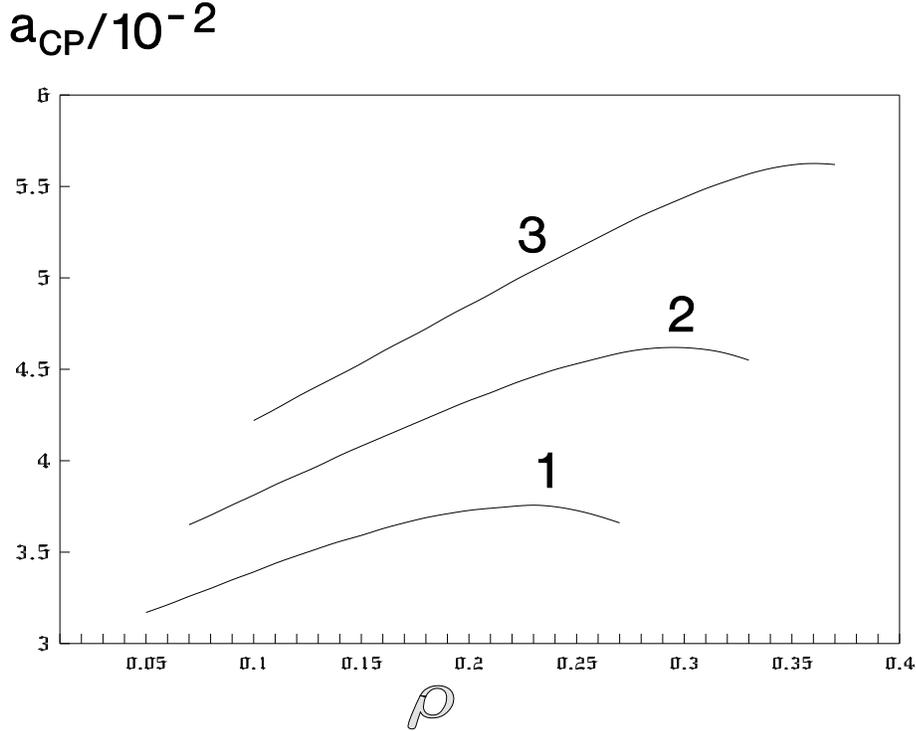}
\caption{The dependence of $a_{CP}(B \to X_d e^+ e^-)$ on CKM parameters
for $\rho^2 + \eta^2 = 0.36^2$; $0.42^2$; $0.48^2$
(lines 1,2,3 respectively).
Other relevant SM parameters are taken at their central values,
the low-energy and matching scales are chosen as $\mu=m_b$,
$\mu_W=M_W$.}
\end{center}
\end{figure}

The predictions of the Models~I~and~II coincide (within $(10 \div 15)\%$
accuracy) with those of the Standard Model. For the Model~III the derived
interval of $B \to X_d e^+ e^-$ rate asymmetry for different choices of 
the low-energy and matching
scales is presented in Table~4.2 . 
As one can see from this table, except
of the case when $\mu=m_b/2$ and $\mu_W=m_t$, the direct CP-asymmetry in
$B \to X_d e^+ e^-$ decay in the 2HDM does not exceed the SM
maximum value more than $(10 \div 15)\%$. On the contrary, the minimum value
of $a_{CP}(B \to X_d e^+ e^-)$ can deviate from the SM result
significantly: it can be $1.5 \div 2.5$ times smaller
than in the SM depending on the choice of the low-energy and
matching scales. $B \to X_d e^+ e^-$ decay asymmetry can also change 
the sign, as compared to that
in the SM. However in this case it 
remains small in absolute value and hence unmeasurable.

Thus our main result in this section is that the 2HDM effects in general
decrease the direct
CP-asymmetry in $B \to X_d e^+ e^-$ decay as compared to 
the predictions of the SM. 
On the other hand, in the SM for
the appropriate choice of the low-energy and matching scales,
$B \to X_d e^+ e^-$ decay asymmetry can be driven to be arbitrary small
(in absolute value). It remains small also in the 2HDM. As a result,
the SM and 2HDM intervals of the CP-asymmetry do not differ
significantly: one obtains $a_{CP}(B \to X_d e^+ e^-)= (0.2 \div 9.8)\%$
in the  SM and $a_{CP}(B \to X_d e^+ e^-)= (-0.5 \div 10.7)\%$ for the 2HDM.
Thus when calculating
$B \to X_d e^+ e^-$ decay asymmetry in the lowest non-vanishing order
one is not able to conclude if this quantity is efficient for testing the
physics beyond the Standard Model. To clarify this point  
one should include  higher-order corrections to
$a_{CP}(B \to X_d e^+ e^ - )$.

\section{Summary and Conclusions.}
We have considered the direct CP-asymmetry in the inclusive $B\to X_d\gamma$
and $B\to X_d e^+e^-$ decays in the framework of general extension
of the Standard Model with two Higgs doublets. The investigation has
been performed in the lowest non-vanishing order in the 
renormalization group improved perturbation theory
taking into account the restrictions on the parameters of the model
coming from the assumption of the absence of New Physics between the
electroweak breaking and unification scales as well as the
experimental constraints from the measurements of the $B\to X_s\gamma$
branching ratio and $B-{\bar B}$ mixing.

It has been shown that in the most general version of the 2HDM (Model~III)
large deviations for $a_{CP}(B\to X_d\gamma)$ from its SM prediction
are possible. In the case of the absence of a new CP-violating phase
(additional to the CKM one) we have found that $a_{CP}(B\to X_d\gamma)$
can be both negative and positive varying in the ranges $(-24\div-3)\%$
and $(10\div 36)\%$ respectively.
The measurement of the sign of $a_{CP}(B\to X_d\gamma)$ can determine
the sign of the Wilson coefficient $C_7^{(0)}(\mu)$. When an additional
source of CP-violation is present our prediction for $a_{CP}(B\to X_d\gamma)$
is $(-25\div 37)\%$. So the only new effect connected with
the Higgs vacuum phase is that $|a_{CP}(B\to X_d\gamma)|$ can be
arbitrary small and hence unmeasurable.

Our results suffer from the large dependence on the low-energy ($\mu\sim m_b$)
and matching ($\mu_W\sim M_W, m_t$) scales when the contribution of the
Higgs boson mediated loops is sizable. For instance, the minimum
value of $a_{CP}(B\to X_d\gamma)$ varies from $-25\%$ to $-7\%$
depending on the values of the scales.

The problem of the renormalization scale dependence is actual also for
$B \to X_d e^+ e^-$ decay asymmetry. Here the situation is
more dramatic: though for some choices of the low-energy scale
New Physics contribution decreases
$a_{CP}(B \to X_d e^+ e^-)$ significantly, 
its allowed interval does not differ
essentially from that in the SM. In other words the calculations
performed in the lowest non-vanishing order of the perturbation theory
do not allow to conclude if the $B \to X_d e^+ e^-$ decay asymmetry
is efficient for testing New Physics beyond the Standard Model.
We expect that the problem of the renormalization scale dependence 
will be resolved after taking into account $O(\alpha_s^2)$ 
corrections
to $a_{CP}(B \to X_d \gamma)$ and
$O(\alpha_s)$ corrections to $a_{CP}(B \to X_d e^+ e^-)$. 

One of the authors (H.M.A.) is indebted to the National 
Research Center Demokritos for kind hospitality. The work
of H.M.A. was supported in part by the Hellenic Ministry of 
National Economy Fellowship Grant  and of G.K.S. by 
EEC Grant no. HPRN-CT-1999-00161. The work of H.H.A and G.K.Y.
was partially supported by INTAS under contract INTAS-96-155.

\end{document}